\documentclass{aa} 
\usepackage{graphicx}
\begin{document}        


\def\ra{\rightarrow} \def\deg{\ifmmode^\circ\else$^\circ$\fi}
\def\kms{km\thinspace s$^{-1}$}
\def\msun{M$_{\odot}$}
\def\lsun{L$_{\odot}$}
\def\h13co+{H$^{13}$CO$^+$}


\title{The structure of Onsala\,1 star forming region} 
\subtitle{}

\author{ M.  S.  N. Kumar\inst{1}, M. Tafalla\inst{2}
\and R. Bachiller\inst{2}}

\institute{Centro de Astrof\'{\i}sica da Universidade do Porto, Rua
das Estrelas, 7150-462 Porto, Portugal\\ e-mail:nanda@astro.up.pt,
\and Observatorio Astron\'omico Nacional, Alfonso XII, 3, E-28014\,Madrid, Spain
}

\offprints{M.  S.  Nanda Kumar} 
\date{Received:  ; accepted:  }
\authorrunning{Kumar et al.}  
\titlerunning{Onsala 1}

\abstract{
We present  new high-sensitivity high-resolution  mm-wave observations
of the  Onsala\,1 ultra-compact  HII  region that  bring to light  the
internal structure of this   massive star forming cloud.   The 1.2\,mm
continuum    map  obtained    with   the  IRAM    30-m  radiotelescope
($\sim$11$\arcsec$ resolution) shows  a  centrally peaked condensation
of 1{\arcmin} size ($\sim$0.5\,pc at the assumed distance of 1.8\,kpc)
which has been further investigated at higher  resolution in the 3\,mm
continuum and  in the emission lines of  \h13co+ J=1--0 and SiO J=2--1
with the IRAM Plateau de Bure interferometer.   The 3\,mm data, with a
resolution  of $\sim$5$\arcsec\times$4$\arcsec$, displays a unresolved
continuum source   at the peak  of  the extended 1.2\,mm  emission and
closely associated with the ultra-compact HII region.  The \h13co+ map
traces the central condensation in good agreement with previous NH$_3$
maps of Zheng et al.\cite{zheng85}.   However,  the velocity field  of
this central condensation, which was previously thought  to arise in a
rapidly rotating structure, is better explained in  terms of the dense
and  compact component of  a  bipolar outflow.  This interpretation is
confirmed by SiO and CO observations of the full  region. In fact, our
new SiO data unveils the presence of multiple (at least 4) outflows in
the region.  In particular,  there is an  important center  of outflow
activity in the region at about  1$\arcmin$ north of the UCHII region.
Indeed the different outflows are  related to different members of the
Onsala\,1  cluster.   The data presented   here support  a scenario in
which the phases  of massive star  formation begin  much later  in the
evolution  of a cluster and/or UCHII  region last for much longer than
10$^5$yrs.

}

\maketitle

\section{Introduction}

Onsala\,1 (ON\,1) is an  ultra-compact HII (UCHII) region  situated in
the densest part of the Onsala molecular cloud (Israel \& Wootten
\cite{iw83}).  It  is  one of  the  most  compact  and isolated  UCHII
regions known to be associated  with several signposts of massive star
formation  such  as dense  gas,    high velocity  gas,  masers,    and
near-infrared  sources.  However, a  single luminous (10$^4.2$\,\lsun)
far-infrared  (FIR)  point source  namely IRAS20081+3132 is associated
with this UCHII region. The luminosity estimated from FIR data and
the  centimeter(radio) emission  agree   well,  within  a  factor   of
100\,\lsun, and indicate a spectral type of  B0.3-B0.6 (MacLeod et al.
\cite{macleod98}).        At        a      spatial  resolution     of
0.7\arcsec-1.5\arcsec,  the UCHII is  detected as an unresolved source
at 2\,cm \& 3\,cm  wavelengths  (Kurtz et al.\cite{kurtz94}).   Higher
angular  resolution(0.1\arcsec)  observations at  1.2\,cm resolve  the
UCHII region into an  approximate  shell structure of size  0.6\arcsec
(Turner \& Matthews \cite{tm84}).  This source is placed at a distance
of 1.3-6\,kpc  by  various authors  in   the  literature.  However,  a
distance of less than 2\,kpc is favored by most authors for many valid
reasons (see e.g.  MacLeod et al.\cite{macleod98}; Watson et al.
\cite{wat03}).  We therefore   adopt the near-distance of  1.8\,kpc to
this source for the purposes of this work.

Several masers such  as OH (Ho et al.  \cite{ho83}), H$_2$O (Downes et
al.  \cite{dow79}), CH$_3$OH (Menten  \cite{men91}) and HCHO  (MacLeod
et al.  \cite{macleod98}) are known to coincide with ON\,1.  Spatially
separated  blue and red shifted OH  maser spots around ON\,1 have been
identified by Desmurs  \& Baudry (\cite{db98}) using  VLBI techniques.
Observations using  dense gas and dust tracers  have  shown a compact,
nearly   circular dense  core surrounding  ON\,1.    Maps of 350$\mu$m
continuum emission (Mueller   et al.  \cite{mueller00}) and   CS J=5-4
emission  (Shirley   et  al.   \cite{shir03})  show    dense  circular
condensations  around  the  source.   Emission   from transitions   of
CH$_3$CN J=12-11 (Pankonin  et al.  \cite{pank01}) and HNCO (Zinchenko
et al.  \cite{zinch00})  are also found to  peak on  ON\,1.  Using VLA
observations at    11\arcsec  resolution in   the (J,K)=(1,1)  line of
NH$_3$, Zheng et  al.   (\cite{zheng85}, hereafter ZHRS85) argued  for
the presence of a rapidly rotating gas condensation centered on ON\,1.
However,   the  regions of massive   star  formation often  possess an
intricate kinematic  behavior involving motions  of rotation,  infall,
and outflow, and the actual  existence of a  rapid gas condensation in
ON\,1 needed confirmation with higher resolution observations. In this
paper,   we present high-sensitivity high-resolution  ($\sim$5\arcsec)
line and continuum observations    which are used to investigate    in
detail the dense core and outflowing gas associated with ON\,1.

\section{Observational Data}

The IRAM 30\,m radiotelescope at Pico Veleta (near Granada, Spain) was
used to map the $\lambda$\,1.2\,mm  continuum emission from the  ON\,1
region   in December 2000.  We used   the  MAMBO\,1 bolometer array to
produce an on-the-fly map  with a scanning  speed of $4''$ s$^{-1}$, a
wobbler period of  0.5 s, and a wobbler  throw of  $53''$. The central
frequency of the   detectors  was $\sim$240\,GHz,  and  the  bandwidth
$\sim$70\,GHz.  Sky dips  made  immediately before and  after  the map
were  used  to correct  for the  atmospheric extinction,  and a global
calibration factor of 15000 counts per Jy (based  on an observation of
Uranus) was applied to the data.  The resulting map (produced with the
IRAM NIC software) has an angular resolution of $11''$.

High   resolution interferometric observations    of ON\,1 region were
carried out using the IRAM Plateau  de Bure (PdB) interferometer (near
Grenoble,  France)   in  its  D   (5  antennas)  and  C  (6  antennas)
configurations.  Observations using  the compact D array configuration
were carried out on the nights of 28 May 2002 and 20 July 2002 for two
adjacent  fields;    one  field  centered   on  the   ON\,1  FIR point
source/UCHII region and another at offset 1\arcmin north.  The C array
configuration observations  for  the same fields  were obtained during
the nights of 9  and 10 January  2003.  Each field was  a mosaic  of 9
pointings arranged in   a   3$\times$3  matrix   with a  spacing    of
12\arcsec\, which is roughly equal to half  the primary beam size.  We
used the SIS heterodyne  receivers both at  87\,GHz and 230\,GHz.  The
230\,GHz data was mostly  unusable  due to summer weather  conditions.
Six  spectral  correlators were used to   select three spectral lines,
namely,  SiO J=2--1($\nu$=86.846891\,GHz), H$^{13}$CO$^+$       J=1--0
($\nu$=86.754330\,GHz)  and  CO J=2--1  ($\nu$=230.538\,GHz)  and  two
continuum bands centered at 1.2\,mm and 3\,mm wavelengths.  The system
temperature was  typically 100\,K  at  3\,mm and $>$400\,K at  1.2\,mm
with    a   phase  noise of       $\sim$25\deg-35\deg  at 3\,mm    and
$>$40\deg-50\deg\, at 1.2\,mm.  The D configuration provided baselines
of 24m-82m  while the C  configuration resulted  in baseline  range of
48m-229m.  We used the sources  2023+336 and  2013+370 as the  primary
phase calibrators and MWC349, 2013+370 and 3C345 for flux calibration.
Synthesized         clean     beams         at     3\,mm      measured
$\sim$5.35$\arcsec\times$3.7$\arcsec$  (PA=56\deg) which represent the
typical  resolution of  the interferometric  data  presented here. The
resulting maps does  not include  any short  spacing data from  single
dish observations and thus filter out all large scale structures.

\begin{figure} 
\resizebox{\hsize}{!}{\includegraphics{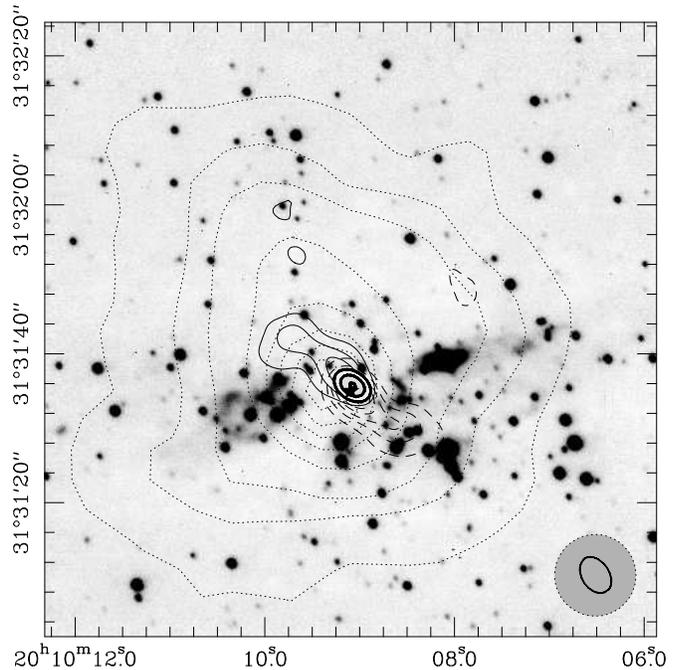}}

\caption{1.2\,mm continuum emission (dotted contours) mapped with 
IRAM30\,m  telescope   overlayed on  a  K-band   image  (Kumar et  al.
\cite{kum03})   of the    Onsala\,1  region.   Contour    levels   are
1,2,3,6,9,12,15,18 \& 21 Jy/beam and has an rms of 140mJy/beam.  Thick
contours enclosing grey shades represent  the 3\,mm continuum emission
mapped with the IRAM PdB interferometer.  Levels are  0.1, 0.15 \& 0.2
Jy/beam (rms = 1.6mJy/beam).  The star  symbol represents the position
of  the  FIR sources  and  the  UCHII region.   Solid  lines represent
blue-shifted and dashed   lines represent  red-shifted \h13co+  J=1--0
emission,        integrated        in       the     velocity     range
V$_{lsr}\pm$4.5\,km\,s$^{-1}$.  In  both   cases, the   contour levels
start at 0.5\,K\,km\,s$^{-1}$  (rms  =  0.06\,K\kms) and increase   in
steps of 0.5\,K\,km\,s$^{-1}$. The beam sizes of the 1.2mm data(dotted
circle) and the  3mm data  (bold  ellipse) are shown enclosed  in grey
shades at the bottom right corner of the figure. }

\label{fig:1} 
\end{figure}

\section{Results}

\subsection{Continuum observations}
In Fig.1  we depict  the  results of  observations that represent  the
dense gas and dust in  the region of ON\,1.  The star symbol shows the
coinciding  position of the  UCHII  region,  the luminous  IRAS  point
source and   all  the maser  spots that  represent   the massive young
stellar object, which  we  shall henceforth  refer to  as  ON\,1.  The
1.2\,mm   continuum  emission  is extended     over a  size  scale  of
$\sim$70$\arcsec$ and  elongated  in  the North-East  direction.   The
1.2\,mm continuum emission is a sensitive tracer of  the warm dust and
therefore  indicates the  dense cloud surrounding  ON\,1.  The spatial
extent and morphology of the 1.2\,mm continuum map  is very similar to
that  of   350$\mu$m map (Mueller  et  al.   \cite{mueller00}), the CS
emission line  maps   (Shirley et al.    \cite{shir03})  and also  the
integrated  NH$_3$ emission  (ZHRS85)  which    has been  thought   to
represent  a rapidly rotating condensation.  The  source ON\,1 and the
surrounding embedded cluster all  appear well immersed in the observed
1.2\,mm emission.  Therefore the 1.2\,mm  emission region in  Fig.\,1,
which is representative   of  the condensation unveiled  by  all other
dense gas tracers, can  be considered as the  dense core in  which the
ON\,1  cluster has been  forming. The 3\,mm continuum emission reveals
an unresolved    source    centered on   ON\,1.    The interferometric
observations filter out the  extended emission and show the underlying
small  scale  structure.    Therefore  the 3\,mm   continuum  emission
centered on ON\,1 is indicative of the  major condensations within the
more extended clump traced by the 1.2\,mm emission.  However, 10$\mu$m
observations (Kumar et al.   \cite{kum03}) reveal two closely situated
sources  coinciding with the  unresolved 3\,mm  source indicating that
ON\,1 is likely to harbor a binary protostar.

The  1.2\,mm emission  maps can be  used  to estimate the mass  of the
ON\,1  clump and   the 3\,mm  continuum   emission  the  mass  of  the
unresolved central source(s).  The black body fits to the Far-Infrared
(FIR) flux density distribution of  ON\,1 indicates a dust temperature
of 57\,K (Mueller et al. \cite{mueller00}). Since the peak of the dust
condensation  coincides with an UCHII  region, this value is likely an
upper  limit.    If we assume    a dust   absorption  coefficient   of
$\kappa_{1.2mm}$=0.0005\,kg$^{-1}$m$^2$       (Henning     et      al.
\cite{henning95})   and a distance  of  1.8\,kpc  to  the source,  the
observed  1.2\,mm  continuum  emission  represents  a   cloud mass  of
$\sim$3860\msun \, for  T=50K and 5000\msun  \, for T=40K.  The source
averaged flux  density (3.6\,Jy/beam) imply hydrogen  column densities
of  1.4$\times$10$^{24}$cm$^{-2}$ resulting in A$_v\sim$700\,mag.  The
peak  column density is  even higher  at 8.2$\times$10$^{24}$cm$^{-2}$
implying a  few thousand  magnitudes of  visual  extinction.  However,
these numbers  are to be  treated with caution due  to the presence of
the UCHII region. The observed  emission does not faithfully represent
the temperature of   the warm dust since the   heating effects of  the
UCHII region can contribute  to a majority   of emission close  to the
peak.  The 3\,mm emission from the unresolved source represents a mass
of          $\sim$400\msun,     where        we      have         used
$\kappa_{3mm}$=6.6$\times$10$^{-5}$\,kg$^{-1}$m$^2$.   This value   of
$\kappa_{3mm}$   is obtained by  assuming a  spectral emissivity index
$\beta$=2.  There  is increasing  observational evidence which suggest
that the  value of $\beta$ is  likely much lower ($\beta\sim$1) around
massive protostellar  sources (ex: Hunter  et al.\cite{hun00}) . If we
assume $\beta\sim$1 then the  resulting estimates of the embedded mass
in the observed 3\,mm emission  will be lower  than quoted above.   In
any case,  the derived mass  from  3\,mm emission is  probably an over
estimate since 3\,mm continuum   emission is known to  trace free-free
emission also and may not  be a good  tracer of the dust alone (Downes
et al. \cite{downes92}).  In view of these facts  we can safely assume
that the enclosed mass in the unresolved central object is not greater
than $\sim$300-400\msun.  Thus about  6-8\% of the overall  dense core
mass seems to be in the form of protostellar material at the center.

\subsection{H$^{13}$CO$^+$ J=1--0 Emission}

In   Fig.~1   solid and dashed   contours   show blue and  red shifted
H$^{13}$CO$^+$  J=1--0  emission,   respectively, integrated  in   the
velocity   range V$_{lsr}\pm$4.5\,km\,s$^{-1}$.   As evident  from the
figure, the  H$^{13}$CO$^+$ emission is centered  on the  source ON\,1
and  reveals  a spatially  resolved bipolar  structure.   The position
angle (PA) of  the axis of this  well-collimated bipolar  structure is
$\sim$44$\deg$ and roughly coincides with  the elongation displayed by
the 1.2\,mm contours.  The H$^{13}$CO$^+$ emission also coincides well
with the NH$_3$ condensation  mapped by ZHRS85.  The projected  length
of  this bipolar emission  is $\sim$30$\arcsec$ corresponding to about
0.27\,pc at   the assumed  distance of  1.8\,kpc.  Each lobe  of  this
compact feature appears to consist of two distinct clumps.

In  Fig.\,2 we show  a velocity-position  map obtained  along the axis
(PA$\sim$44$\deg$) of this  bipolar feature which reveals  a structure
that   is  distinctly different from    earlier claims of ZHRS85.  The
vertical  line   represents     the  systemic   LSR    velocity     at
12\,km\,s$^{-1}$.  The  red-shifted southern lobe and the blue-shifted
northern lobe can be seen clearly separated into two components on the
PV diagram.  A main component  of the emission (denoted by  A \& B  in
Fig.\,2) stretches to   a radius of  $\sim$5$\arcsec$ and  a satellite
component comprising of two  knots (denoted by  C  \& D) extends to  a
radius of $\sim$10$\arcsec$.  These  4 clumps correspond to the clumps
found on the  images (Fig.1).  As can be  seen from Fig.2, all  of the
found emission is spread along a straight line on the PV-plane showing
increasing velocity with increasing distance from the central position
following  a Hubble law.  We have  not subtracted the bright continuum
emission from this data.  The continuum emission can therefore be seen
close  to  the source  (offset  = 0) extending on  either  side of the
V$_{lsr}$,  up   to the    limits  of our   observing  setup  with the
spectrometers.

Since the   H$^{13}$CO$^+$  J=1--0  emission  is symmetrical   both in
position  and  velocity with   respect  to the  central  position  and
velocity,  it is natural to conclude   that such emission  arises in a
bipolar  outflow emerging  from  the  central object.  Note  that  the
observed kinematical  structure can  not  be easily explained  with  a
rotating disk.   First of   all,  the  total  size  of this  structure
($>$0.2\,pc)  is too large to  be a disk  and,  moreover, a disk would
produce just two  peaks of emission in the  PV diagram, not four peaks
as the A,B,C, \& D peaks which are actually  observed in Fig.\,2.  The
Hubble-like velocity law  distribution which  is  observed in  Fig.\,2
(velocity proportional to the distance) is  also very typical of young
bipolar  outflows.  A straight line fit  through all the knots shows a
gradient of $\sim$30\,km\,s$^{-1}$\,pc$^{-1}$  at the assumed distance
of 1.8\,kpc.   These velocity limits   and the velocity  gradient  are
similar  to the  values  obtained from  NH$_3$ observations by  ZHRS85
except   that   the  gradient  of    $\sim$11\,km\,s$^{-1}$\,pc$^{-1}$
estimated by ZHRS85 is due to  their adopted distance of 3.5\,kpc. The
interpretation that the H$^{13}$CO$^+$  emission arises in the compact
component of  a bipolar outflow  is confirmed by the SiO high-velocity
data shown in the following section.

\begin{figure} 
\resizebox{\hsize}{!}{\includegraphics{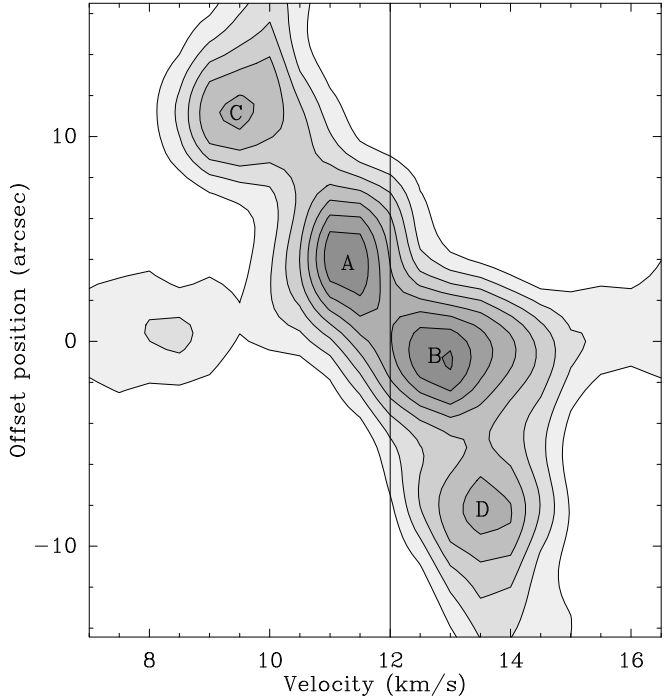}}

\caption{Velocity-Position diagram for the H$^{13}$CO$^+$ J=1--0 
emission  shown  in Fig.\,1, obtained along   the axis of  the bipolar
emission at PA$\sim$44$\deg$. The contours  levels start at 1\,K
and increase in steps of 1\,K. }
\label{fig:2} 
\end{figure}

Assuming standard LTE conditions,  the observed \h13co+ emission can be
used to compute the column densities and enclosed mass. If we assume a
dipole moment of  $\mu$=4.08\,debye (Haese \& Woods\cite{hw79}) and an
H$^{13}$CO$^+$/H$_2$  ratio  of  3.3$\times$10$^{-11}$   (Blake et al.
\cite{blake87}; in OMC1), the   observed fluxes indicate an  enclosed
mass of $\sim$280\msun-330\msun for  temperatures of 40\,K-50\,K.  The
estimated mass can   be  uncertain  by  a  few  hundred solar   masses
depending on the assumed  value of \h13co+/H$_2$ ratio  and excitation
temperature.  In any case, the estimated  \h13co+ mass  is of the same
order than the mass evaluated from the 3\,mm continuum emission.

\begin{figure} 
\resizebox{\hsize}{!}{\includegraphics{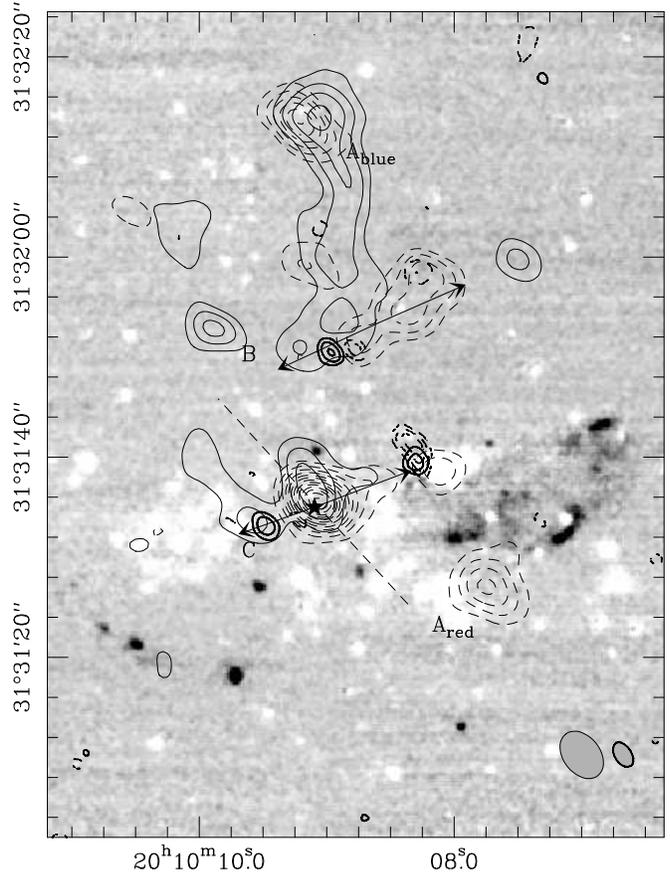}}

\caption{SiO J=2--1 (thin lines) and CO J=2--1 (thick lines) integrated  
emission overlayed  on a continuum subtracted  H$_2$ narrow band image
at 2.122$\mu$m.  SiO emission  was integrated with a velocity interval
of V$_{lsr}\pm$4\,km\,s$^{-1}$ and  CO emission was  integrated in the
range V$_{lsr}\pm$12\,km\,s$^{-1}$ Solid lines represent  blue-shifted
and dotted  lines represent red-shifted  emission.  SiO contour levels
start at  0.2\,K\,km\,s$^{-1}$ (rms  =  0.06\,K\kms) and  increase  in
steps    of  0.15\,K\,km\,s$^{-1}$.   CO   contours   levels start  at
2\,K\,km\,s$^{-1}$ (rms =  0.22\,K\,kms)  and  increase in steps    of
1\,K\,km\,s$^{-1}$.  The  star symbol  represents the position  of the
FIR sources and  the UCHII region as in  Fig.\,1.  The dotted straight
line represents the axis of the \h13co+ outflow. The beam sizes of the
3mm and  1mm  data are  shown by filled  ellipses at  the bottom right
corner of the figure.}

\label{fig:3} 
\end{figure}

\subsection{SiO \& CO observations}

In Fig.\,3  we  present the results of  the  SiO J=2--1 and  CO J=2--1
high-resolution observations in all  the region around ON\,1 including
the north zone in which the  1.2\,mm emission is significant.  The SiO
emission is  found at velocities  of V$_{lsr}\pm$4-5\kms, with respect
to the systemic velocity, whereas the CO emission appears to extend up
to V$_{lsr}\pm$12\kms.  The  CO J=2--1 data at  1.2\,mm  is quite poor
due to summer weather conditions, but  the knots shown in Fig.\,3 have
high signal to noise  ratio  (10-15$\sigma$).  Based on  complementary
blue and red  shifted components we  can identify  at least four  main
outflows  in the region.  However, we caution  that  these maps do not
reveal most of the large scale  structures in the  region, since we do
not have any short  spacing data from  single dish telescopes included
in the maps.

The first outflow is  seen close to the central  FIR source.  The most
internal  component  of this outflow  is   at PA$\sim$45$\deg$ in good
agreement   with  the H$^{13}$CO$^+$   outflow  component discussed in
Section  3.2.  As  for  the   H$^{13}$CO$^+$ data,   the  blue-shifted
emission is placed  at  the NE of  the  source, while the  red-shifted
emission  is placed at the  SW.  We then believe  that the SiO and the
H$^{13}$CO$^+$ high velocity   emission  is tracing the   same compact
bipolar outflow  which is emerging from  the very vicinity of  the FIR
object.

Most  remarkably, the region placed  40$\arcsec$ north from ON\,1\,FIR
appears to be  an important center   of outflow activity: a  prominent
high-velocity  SiO feature  which  is marked  as A$_{blue}$ in Fig.\,3
stretches approximately   30$\arcsec$ in   the  north-south  direction
forming a  kind of highly-collimated  jet. the red shifted counterpart
to  this jet-like structure   is not so obvious.    We think  that the
bright  red  knot (A$_{red}$)  on  the  south can   be associated with
A$_{blue}$ because both structures are  well aligned and because there
is no other equally prominent blue  component in the region.  If these
two  features A$_{red}$ and A$_{blue}$ were  actually part of the same
outflow, this  would  have  a  total  size of at  least  1$\arcmin$ or
0.5\,pc.  We caution however that it  is difficult to identify bipolar
counterparts at large scale in such complex regions.

A smaller outflow denoted by $B$ shows both blue and red shifted lobes
and extends to about 25$\arcsec$ in the east-west direction. This flow
is traced even  by the  poor  quality CO  data appearing as  spatially
separated blue and red knots.  The  driving source of this outflow has
not been  identified neither in the  near-infrared nor  in the mm-wave
observations. The  3$\sigma$   noise  limit of   5\,mJy   of our 3\,mm
continuum  maps  imply a    limiting  detectable mass  of  5-6\,\msun.
Therefore this outflow is likely driven by a low mass object.

Finally, a  short   east-west jet-like feature,   denoted  by  $C$, is
emerging from  the dense  blob  of SiO  emission surrounding the ON\,1
source represents another distinct bipolar outflow.  Indeed, this flow
is traced more prominently by the symmetrically  placed CO knots in an
axis  that  closely  matches with the  axis    of the overall  bipolar
distribution of near-infrared H$_2$  emission.  It is unclear if  this
outflow arises from  the same object  than the dense and compact NE-SW
outflow seen  in H$^{13}$CO$^+$ and SiO, and  which has been discussed
above. Since the   FIR object is   probably a binary  system (Kumar et
al.  \cite{kum03}), it  may be  not  surprising to find two  different
outflows emanating from its vicinity.

It appears  anyway that high velocity emission  is spread all over the
cluster forming molecular  cloud.  Although our observations  indicate
the presence of real multiple outflows, it is also likely that some of
these outflows have interacted with the surrounding cluster and broken
into multiple  components.   This may   be particularly  true for  the
east-west flow which is traced  by the CO knots, a  faint SiO jet  and
largely by the NIR H$_2$ emission. 

\section{Discussion}

ON\,1 was described  to be a rapidly  rotating condensation by  ZHRS85
based  on NH$_3$ observations.   The higher  resolution data presented
here along with other dense gas and dust tracing observations indicate
that the NH$_3$  condensation described by ZHRS85 actually corresponds
to an outflow.  The \h13co+  observations discussed in Section 3.2 has
the orientation  and velocity ranges similar  to  the NH$_3$ features.
However, for    the first time, our  observations    shows a spatially
resolved bipolar  feature whose PV   diagram  is typical of  outflows.
Perhaps the  \h13co+ observations  alone  would not have  resolved the
issue if  this  is  an  outflow   or  a  rapidly rotating    envelope,
particularly because of the velocity structure close to the source and
the  low values   of $\Delta  v  =  V_{lsr}\pm$2-3\kms.  But the   SiO
emission which traces other outflows in the  region show only slightly
higher  values of  $\Delta  v =  V_{lsr}\pm$4-5\kms.  The  symmetrical
lobes of  \h13co+ emission  indicate   that the  outflow has  a  small
inclination angle with respect to the  plane of the sky.  This implies
small line-of-sight velocities  of the \h13co+ structure in comparison
to the SiO data.  Indeed all other dense gas  tracers from this region
such  as CS (Shirley et    al.\cite{shir03}),  CH$_3$CN (Pankonin   et
al.\cite{pank01})  and  HNCO (Zinchenko et  al.\cite{zinch00}) display
similar V$_{lsr}\sim$11-12\kms and  $\Delta v \sim$4-5\kms.  Therefore
it is clear  that  the \h13co+  presented here spatially  resolves the
NH$_3$ data from ZHRS85 and reveal an outflow well placed in the plane
of the sky and possibly slow moving.  A similar  example where a large
scale coherent NH$_3$ structure  (Jackson et al. \cite{jackson88})  is
aligned well with an outflow (Bachiller  \& Cernicharo \cite{bc90}) is
found in the case of NGC6334I.

In the light of  above  results, the  following scenario appears  most
plausible    for the ON\,1   region.   The  UCHII   region  ON\,1 is a
relatively isolated massive   star forming  region associated  with  a
small embedded cluster, all of which is immersed  in a molecular clump
of few   thousand solar masses.    The  vicinity of the  UCHII  region
appears to  drive at least two (the  \h13co+ compact outflow, the flow
traced by CO knots and   H$_2$) and likely   three (SiO flow A)  major
outflows indicating  multiple embedded sources of similar evolutionary
states.  While  the exact driving sources  for each of these flows are
not clear, it is very likely that the \h13co+  flow originate from one
of the two mid-infrared (10$\mu$m) sources (Kumar et al. \cite{kum03})
centered on ON\,1.   These two mid-IR sources  are neither visible  at
3.8$\mu$m   nor  distinguishable   by  the   millimeter  observations.
However,  given that  both  sources  show 10$\mu$m  emission and   are
associated   with an unresolved mm  continuum   peak, and likely drive
outflows, they may be at a similar evolutionary state.  Thus the ON\,1
source could harbor a massive binary protostar.

The   \h13co+   and  SiO outflows   display   very  low  line-of sight
velocities. Low values of line of  sight velocities in outflows can be
expected if they lie in the plane of the sky as appears to be case for
\h13co+  flow.  But the  SiO  flow  does  not obviously  appear  to be
oriented  in the plane  of the  sky therefore  indicating that the low
line  of sight velocities  are    representative of the region.    The
outflow traced by   \h13co+ does  not seem  to  show any  obvious  SiO
counterparts.  SiO production requires  violent shocks and dust  grain
destruction   while  \h13co+   outflow  emission  only  requires   the
acceleration of the dense  core material (commonly traced by \h13co+).
Given the high density of the central regions of this symmetric clump,
the outflow traced by \h13co+   is likely encountering viscous  forces
thus  only able to  accelerate  the gas   rather than produce  violent
shocks.  Further,  since all  dense  gas tracers  also display similar
range of  low   velocities  such  as   the observed   outflows, it  is
suggestive that the outflows are experiencing an opposing force in its
passage  through the  very  dense, cluster   forming core.  Since  the
jet/outflow experiences the viscous forces uniformly from the point of
ejection, the force only makes the flow have  much lower velocities in
comparision to other outflows.

The presence  of a surrounding embedded  cluster to ON\,1 UCHII region
represents  a low mass stellar  population that could be anywhere from
Class I to  pre-main sequence phases.  If  we  consider that  low mass
stars   form in  $\sim$10$^6$yrs and       massive  stars form      in
$\sim$10$^5$yrs, the simultaneous presence of two massive outflows and
the central location  of the luminous  sources provide two alternative
scenarios. Either the massive star formation began much later than low
mass star formation in this cluster forming core or that the formation
of massive   star/UCHII  region  phases last   for  much  longer  than
$\sim$10$^5$yrs.  While  Herbig (\cite{herbig62}) suggested  that
massive stars form  {\em preferentially}  late  in the evolution of  a
cluster, Stahler (\cite{stahler85}) argued that it is indeed so due to
statistical  reasons.  In  a    related   context,  de Pree   et   al.
(\cite{depree95}) suggest that the life times of the UCHII regions can
be  much longer than a  few thousand years if they   form and exist in
dense and  warm  environments.  The black  body fits  to  the FIR flux
densities of several  relatively isolated massive star forming regions
(ex:   Hunter et   al.\cite{hun00};  Mueller  et  al.\cite{mueller00};
Beuther et  al.   \cite{beu02a})  suggest   high temperatures  between
50-100K  and  densities $\ge$10$^5$cm$^{-3}$ supporting  the  scenario
predicted by de Pree et al.  (\cite{depree95}).  This is also true for
ON\,1    in  view of  the      high  estimated column  densities   and
temperatures. However these observations  can not effectively rule out
the possibility that massive star  formation initiated much later than
the surrounding cluster of low mass stars.

The FIR and radio flux of the ON\,1 source and the mass estimates from
this work  suggest that ON\,1  is  among  the early  B type  candidate
massive protostars. This source can be compared  to other well studied
massive      protostars    such   as     IRAS05358+3843  (Beuther   et
al. \cite{beu02b}),  IRAS18556+0136 (Fuller et al. \cite{fuller01}) or
W75N (Shepherd et  al.  \cite{shep03}).  While W75N and IRAS18566+0136
have UCHII regions,  IRAS05358+3843 has none and  is  likely about the
same age as the massive protostars such as IRAS20126+4104 (Cesaroni et
al.      \cite{ces99})    and     IRAS23385+6053         (Fontani   et
al.\cite{fontani04}).  Among all  these, ON\,1 is  a neat example of a
circularly symmetric molecular   condensation with a centrally  placed
massive  binary  protostar which  is    driving multiple outflows  and
surrounded by a young stellar cluster.

\begin{acknowledgements}

We gratefully acknowledge Roberto  Neri and Sebastian Muller for their
extremely helpful support     during   the reduction  of     the  IRAM
interferometric   data.  This   work  has  been   supported   by grant
POCTI/1999/FIS/34549 approved by FCT   and POCTI, with funds  from the
European    Community  programme  FEDER,  and  by   Spanish  MEC grant
PNAYA2000-0967.

\end{acknowledgements}


\begin{thebibliography}{}

\bibitem[1990]{bc90}

Bachiller, R. \& Cernicharo, J. 1990, \aap, 239, 276

\bibitem[1987]{blake87}

Blake, G. A., Sutton, E. C., Masson, C. R., \& Phillips, T. G. 1987, \apj, 315, 621

\bibitem[2002]{beu02a}

Beuther,  H., Schilke,   P., Menten,  K.  M.,  Motte,  F.,  Sridharan,
T. K. \& Wyrowski, F. 2002, \apj, 566, 945

\bibitem[2002]{beu02b}

Beuther,  H., Schilke, P., Gueth,   F., McCaughrean, M., Andersen, M.,
Sridharan, T. K., \& Menten, K. M. 2002, \aap, 387, 931

\bibitem[1999]{ces99}

Cesaroni, R., Felli, M., Jenness, T., Neri, R., Olmi, L., Robberto,
M., Testi, L., \& Walmsley, M.  1999, \aap, 345, 949

\bibitem[1995]{depree95}

de Pree, C. G., Rodriguez, L. F. \& Goss, W. M. 1995, RMxAA, 31, 39

\bibitem[1998]{db98}

Desmurs, J. F. \& Baudry, A. 1998, \aap, 340, 521

\bibitem[1979]{dow79}

Downes, D., Genzel, R., Moran, J. M., et al. 1979, \aap, 79, 23

\bibitem[1992]{downes92}

Downes, D., Radford,   S. J. E.,  Guilloteau,  S., Guelin, M.,  Greve,
A. \& Morris, D. 1992, \aap, 262,424

\bibitem[2004]{fontani04}

Fontani, F., Cesaroni,  R., Testi, L.,  Walmsley, C. M., Molinari, S.,
Neri, R., Shepherd, D., Brand, J., Palla, F., \& Zhang, Q. 2004, \aap,
414, 299

\bibitem[2001]{fuller01}

Fuller, G. A., Zijlstra, A. A., \& Williams, S. J. 2001, \apj, 555, L125


\bibitem[1979]{hw79}

Haese, N. N., \& Woods, R. C. 1979, Chem. Phys. Lett., 61, 396

\bibitem[1962]{herbig62}

Herbig, G. H., 1962, ApJ, 135, 736

\bibitem[1985]{henning95}

Henning, Th.,   Michel, B., \&  Stognienko,   R. 1995,  Planet.  Space
Sci. (Special issue: Dust, molecules and backgrounds), 43, 1333

\bibitem[1983]{ho83}

Ho, P.  T. P., Haschick, A. D.,  Vogel, S.  N., \&  Wright,  M. C. H.,
1983, \apj, 265, 295

\bibitem[2000]{hun00}

Hunter, T. R., Churchwell, E., Watson, C., Cox, P., Benford, D. J. \& Roelfsema, P. R. 2000, \aj, 119, 2711

\bibitem[1983]{iw83}

Israel, F. P., \& Wootten, H. A. 1983, \apj, 266, 580

\bibitem[1988]{jackson88}

Jackson, J. M., Ho, P. T. P. \& Haschick, A. D. 1988, \apj, 333L, 73

\bibitem[2002]{kbd02}

Kumar, M.  S.  N., Bachiller, R., \& Davis, C.  J.  2002, \apj, 576,
313 

\bibitem[2003]{kum03}

Kumar, M. S. N., Davis, C. J. \& Bachiller, R.  2003, Ap\&SS, 287,191

\bibitem[1994]{kurtz94}

Kurtz, S., Churchwell, E., \&  Wood, D. O. S. 1994, \apjs, 91, 659


\bibitem[1998]{macleod98}

MacLeod, G. C., Scalise, E., Saedt, S., Galt, J. A., \& Gaylard, M. J. 1998, \aj, 116, 1897

\bibitem[1991]{men91}

Menten, K. 1991, \apj, 380L, 75

\bibitem[2000]{mueller00}

Mueller, K. E., Shirley, Y. L., Evans, N. J., II \& Jacobson, H. R. 2000, \apjs, 143, 469 

\bibitem[2001]{pank01}

Pankonin, V., Churchwell, E., Watson, C. \& Bieging, J. H. 2001, ApJ, 558, 194

\bibitem[2003]{shep03}

Shepherd, D., Testi. L., \& Stark, D. P. 2003, ApJ, 584, 882

\bibitem[2003]{shir03}

Shirley, Y. L., Evans, N. J., Young, K. E., Knez, C. \& Jaffe, D. T. 2003, ApJS, 149, 375 

\bibitem[1985]{stahler85}
Stahler, S. W., 1985, ApJ, 293, 207

\bibitem[1984]{tm84}

Turner, B. E., \& Matthews, H. E., 1984, ApJ, 277, 164

\bibitem[2003]{wat03}

Watson, C., Araya, E., Sewilo, M., Churchwell, E., Hofner, P., \& Kurtz, S. 2003, ApJ, 587, 714

\bibitem[1998]{zhang98}

Zhang, Q., Hunter, T. R., \& Sridharan, T. K., 1998, ApJ, 505, L151

\bibitem[1985]{zheng85}

Zheng, X. W., Ho, P. T. P., Reid, M. J., \& Schneps, M. H. 1985, ApJ, 293, 522

\bibitem[2000]{zinch00}

Zinchenko, I., Henkel, C. \& Mao, R. Q. 2000, \aap, 361, 1079
 
\end{thebibliography}
\end{document}